# The Sun and its Planets as detectors for invisible matter

Sergio Bertolucci [1,2], Konstantin Zioutas [1,3], Sebastian Hofmann [4], Marios Maroudas [3]

1. CERN, Geneva, Switzerland
2. INFN, LNF, Italy
3. University of Patras, Patras, Greece
4. Munich, Germany

Emails: Sergio.Bertolucci@cern.ch ; Konstantin.Zioutas@cern.ch

**Abstract:**

Gravitational lensing of invisible streaming matter towards the Sun with speeds around $10^{-4}$ to $10^{-3}$c could be the explanation of the puzzling solar flares and the unexplained solar emission in the EUV. Assuming that this invisible massive matter has some form of interaction with normal matter and that preferred directions exist in its flow, then one would expect a more pronounced solar activity at certain planetary heliocentric longitudes. This is best demonstrated in the case of the Earth and the two inner planets, considering their relatively short revolution time (365, 225 and 88 days) in comparison to a solar cycle of about 11 years. We have analyzed the solar flares as well as the EUV emission in the periods 1976-2015 and 1999-2015, respectively. The results derived from each data set mutually exclude systematics as the cause of the observed planetary correlations. We observe statistically significant signals when one or more planets have heliocentric longitudes mainly between $230^o$ and $300^o$. We also analyzed daily data of the global ionization degree of the dynamic Earth atmosphere taken in the period 1995-2012. Again here, we observe a correlation between the total atmospheric electron content (TEC) and the orbital position of the inner three planets. Remarkably, the strongest correlation appears with the phase of the Moon. The broad velocity spectrum of the assumed constituents makes it difficult at this stage to identify its source(s) in space. More refined analyses might in the future increase the precision in the determination of the stream(s) direction and possibly allow conclusion of some properties of its constituents.

## 1. Introduction

The detection of the constituents of dark matter is one of the central challenges in modern physics. The strongest evidence of dark matter (DM) comes from large scale gravitational observations, while direct and indirect searches have so far provided no convincing evidence of it. The large scale observations suggest that the ordinary dark matter halo in the Galaxy is rather uniform, at least for the size of the solar system; in literature, the co-existence of dark streams or the galactic dark disk hypothesis have also been considered (see e.g. [1,2]). The existence of such streams of dark matter could explain the somewhat puzzling behavior of the active Sun, where there is not yet a consensus model on the origin of solar energetic phenomena, like the solar flares [3,4] and the unnaturally hot Corona [5,6]. In this work we refer to generic dark candidate constituents as "*invisible massive matter*", in order to distinguish them from ordinary dark matter.

We address here the as yet unanswered intriguing question as to whether the motor of the active Sun is entirely of an internal nature, or if it is triggered by some external influence. We follow the latter scenario, by assuming that the triggering mechanism is the planetary lensing of the invisible massive matter stream(s). This scenario is totally different from the models based on tidal forces, which have been attempted with very little success since the discovery of the first large flare some 155 years ago. Presently, we are making neither assumption about the nature of the streaming *invisible massive matter* nor on its interaction with normal matter in the Sun: our goal is to prove the lensing and the existence of preferred direction(s). If this seminal idea holds, there will be ways to explore it further in the future, due to its implications in other ongoing dark matter searches.

## 2. Streaming dark matter and planetary gravitational lensing

Due to the non-relativistic velocities of DM candidates, planetary gravitational lensing becomes efficient [7]. We recall that the deflection angle (Θ) is given by $\Theta \sim \frac{M(R)}{v^2 R}$, where M(R) is the mass inside the radius R, which is taken to be equal to the impact parameter of the impinging particle moving with velocity $v$. Jupiter can focus streaming DM constituents at the Earth / Sun position with velocities in the range ~$10^{-3}$ – $10^{-2}$c [7]; the Sun can focus anywhere on its planetary system particles with speeds in the range ~$10^{-2}$ – $10^{-1}$c. In addition, the Earth and the Moon have a focal length of ≥1 AU for particles moving with $v \approx 10^{-2}$c and $v \approx 3 \cdot 10^{-3}$c, respectively. In particular, the Moon can focus exotic particles with $v \approx 2 \cdot 10^{-4}$c onto the Earth.

Therefore, it is reasonable to assume that constituents of any kind of *invisible massive matter* having a wide velocity spectrum around 300 km/s can undergo gravitational focusing towards the Sun, if one of the planets considered in this work, i.e., Mercury, Venus and/or Earth, enters inside such a stream towards the Sun. A planetary gravitational lens can enhance the flux of invisible particles into the Sun (or other places of the solar system) by up to a factor of ~$10^6$ [7], if a stream crosses the ecliptic plane. The focusing efficiency becomes maximum, if an invisible stream is collinear (*ideally*, within about 0.1°) with the line connecting one planet and the Sun. The existence of invisible streams has been put forward in many studies such as the one for relic neutrinos ([8], see also [2]).

The existence of cosmic streams of normal matter is proven in the case of the neutral Helium, which flows with a velocity peaking at about 25 km/s (see, e.g., ref. [9]). This velocity range is then within the focusing efficiency of the solar system.

**3. Externally driven solar activity: the concept**

In this work we provide primarily a statistical analysis of time (=position) distributions of the 3 inner planets in association with the occurrence of M-class and the X-class flares in the period 1976-2015. We also analyze the continuous solar emission in the EUV ($E_\gamma >$ 24eV), in the period (1999-2015), and the electron content of the Earth ionosphere in the period 1995-2012. The driving idea behind this analysis is based on the following assumptions:

a. Slow moving invisible (streaming) matter of galactic / cosmic origin, whatever its eventual properties, interacts *somehow* with the Sun. The current missing signature in direct searches for dark matter particles such as axions or WIMPs is not necessarily in contradiction with this assumption. This is because their extremely feeble interaction with ordinary detector material excludes them from being viable candidates for this work. For example, dark matter axions require fine tuning of a resonance cavity inside a magnetic field for their conversion to a real photon to happen, and relevant experiments have not reached the necessary sensitivity yet. Similarly, for energetic axions or axion-like particles to oscillate to photons, the required resonance condition, $m_{axion}=m_{\gamma'}$, may be established in the solar atmosphere only locally. In the case of WIMPs, ~sub-keV recoil energy threshold effects may disfavour man-made (underground) detectors [2]. For this reason they are still blind to low mass WIMPs with a steeply increasing cross section. For the solar atmosphere the threshold should be in the ~10eV range (for ionisation or atomic transitions to happen), or even much lower for the ionised solar plasma below the Transition Region. The solar atmosphere, which earth bound or space telescopes observe continuously, is after all a unique windowless magnetised gaseous target being sensitive to much lower energy deposition. To our knowledge, in dark matter research a gaseous detector with variable density inside a magnetic field does not exist. In addition, possible screening effects for the unknown invisible constituents are absent only for the solar atmosphere.

b. The stream constituents have a velocity distribution, which will allow planetary gravitational lensing increasing the flux impinging into the Sun. This temporally increased influx may trigger solar activity.
The different planets according to their mass and distance from the Sun select different velocity ranges for an optimal focusing on the Sun. For this reason, we expect, instead of simple synodic alignments, a different planetary configuration to be associated with (transient) solar phenomena. For example, the t.o.f. for 1AU of particles with $v\sim 10^{-3}c$ is 5 days. If the Earth participates with the gravitational lensing, then it will appear some 5° advanced in heliocentric longitude when the solar activity is being triggered. Therefore, we do not expect simple synodic alignments as is the case for particles with $v\approx c$.

c. The analysis of the frequency of occurrence of flares and/or the amplitude of the solar EUV emission and the electron content of the ionosphere is performed as a function of the position of the Earth and of the inner two planets along their orbits. Due to their different revolution times the appearance of an excess around the same longitudinal range will signal the presence of one or more streams, since each planet is passing through the same longitude at different times.

## 4. THE SOLAR OBSERVATIONS: the flaring Sun and its EUV emission.

The distribution in time of M-class solar Flares and the light curve of full disk solar EUV irradiance is shown in Figure 1. In this analysis, planetary positions at a certain date were derived from a NASA program [10] with a binning of 1 day. Both solar observations and planetary positions are available with a finer resolution, which might be used in future investigations.

The M- and X-flare energy threshold is about 100 to 1000 times above the non-flaring Sun level. The quiescent time noise is constant within less than 10% variation [13]. The choice of using the M- and X-class flares avoids any significant noise related effects.

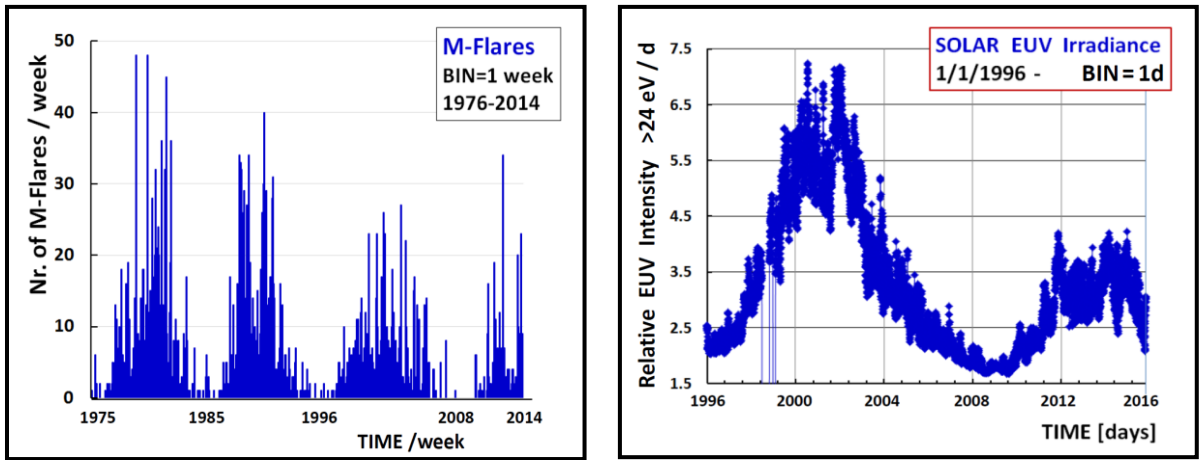

**Figure 1.** (*left*) Number of M-class Flares per week during the period 15/11/1975 - 18/4/2014 [11]. The total number of M-Flares is equal to 6091. (*right*) Light curve of solar emission in the EUV ($E_\gamma > 24$ eV) since 1/1/1996 [12]. The vertical thin lines indicate the intervals without data. In this work we use data continuously taken from 3$^{rd}$ February 1999.

### 4.1 M- and X-class Flares

### 4.1.1 The data and their quality

We have analyzed 6091 M-class and 491 X-class flares recorded between 1976 and 2015 by the GOES mission [11]. The observations were made by a series of geosynchronous satellites (GOES), which overlap in time. They are in circular orbits at 35,700 km tracking the Sun with full-time coverage providing whole-Sun X-ray fluxes with an X-ray threshold energy at about 1.5 keV. The intrinsic time resolution is equal to 3 s. GOES probably has the highest duty cycle, >(94±4)%. From all solar-dedicated space missions it offers the most complete record of solar flares over the last solar cycles [13], and the lists have been widely used to investigate flare statistics.

There is a degree of undersampling mainly weak flares, because the flux contrast for small flares on top of a light curve of a large flare is relatively weaker, compared with the background noise (quiescent times). The double flares can amount to ~ 5% [14], and there is no reason for them to be correlated to the position of the planets.

### 4.1.2 Data analysis

The Earth's heliocentric longitudinal distribution of the M- and X-class Flares is shown in Figure 2. It is difficult to infer only from one plot a particularly significant clustering. Moreover, and whatever clustering is observed, a single planet will never be able to distinguish between a single inner solar clock mechanism from an external cause. Figure 3 shows a similar procedure applied to Venus. As in the case of the Earth, we have not corrected the expected modulation in orbit (stay time) due to the small eccentricity of its orbit. A strong and wide peak around ($260°\pm10°$) heliocentric longitudes appears for both M- and X-Flares as well as a second one at ~145°. We note that the enhancement around 260° is happening around the same heliocentric longitude as seen for the Earth (~255° in Figure 2), despite the fact that Venus and Earth have gone through the same longitude mostly at different times. Figures 4-5 show the corresponding distributions for the much faster Mercury ($T_{orbit}$ = 87.969 days).

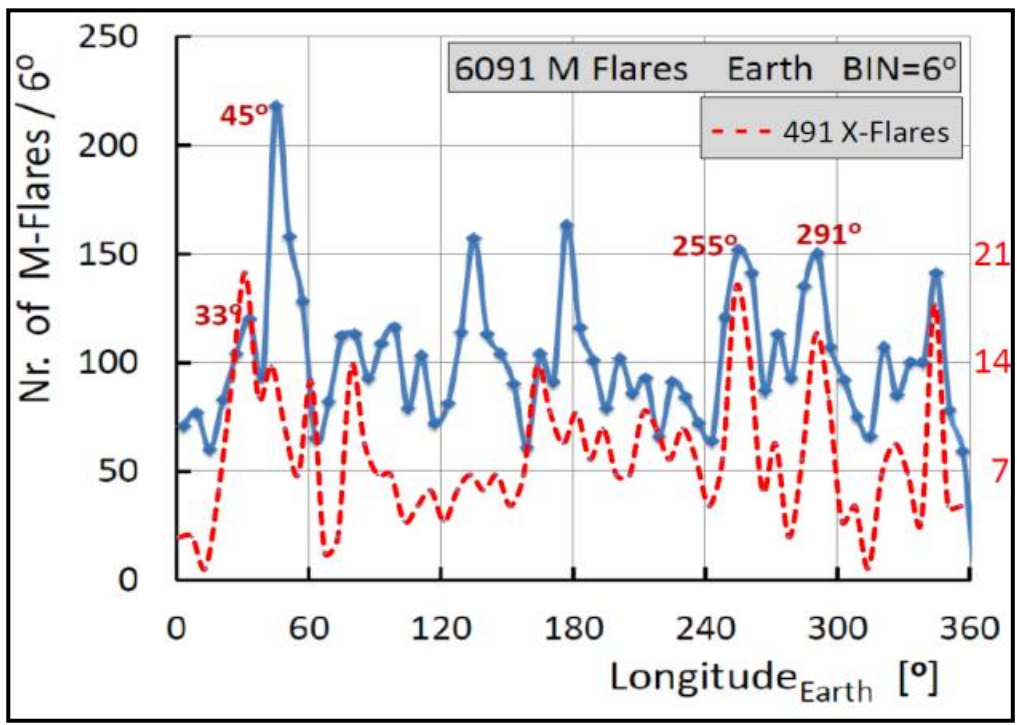

**Figure 2.** Number of M-class Flares (blue line) from the period 15/11/1975 - 18/4/2014 and X-class flares (red dashed line) from the period 1/1/1976 – 5/5/2015 as a function of the **Earths** heliocentric longitude. BIN=6°.

Figure 4a shows a strong modulation, which comes from Mercury's large eccentricity (i.e., the stay-time per longitude-BIN varies in the course of one orbit by a factor of 2.2). The red dashed line shows the effect of the orbit eccentricity in the null hypothesis (i.e. a uniform distribution of flares as a function of time). By subtracting such simulated values (red dashed line) from the raw data (Figure 4a), we obtain Figure 4b, which shows statistically significant excesses at three different longitudes. What is important to stress in our case, is again the appearance of a large excess in the region 240° – 300°, which corresponds to what we have observed for Earth and Venus. In accordance with the working hypothesis of this work, we consider such clear residuals along with their spectral shape as the manifestation of planetary

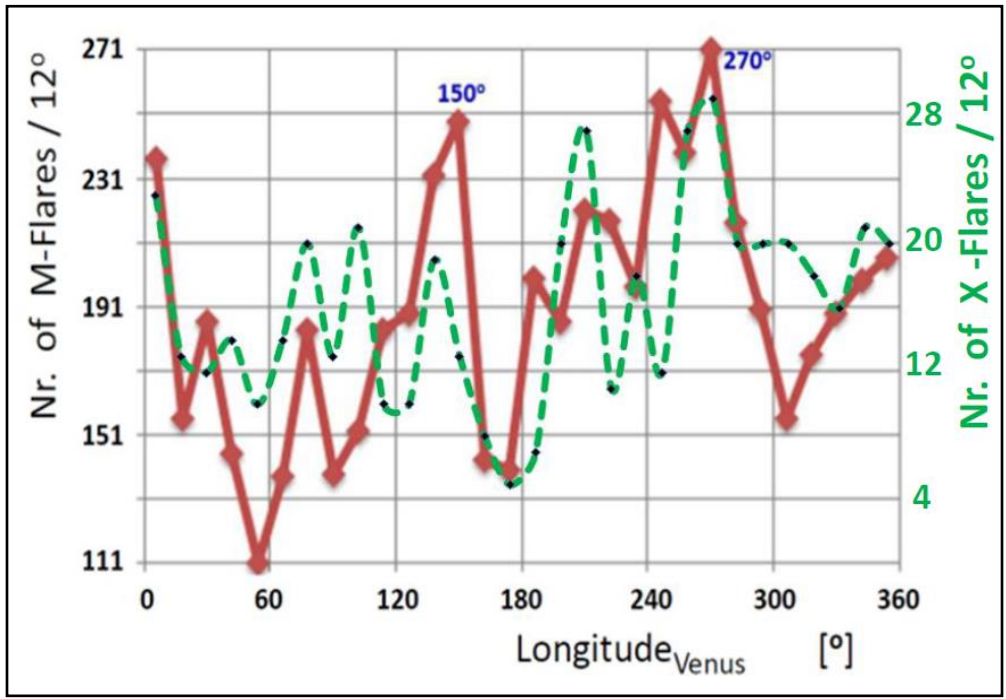

**Figure 3.** Number of M-class Flares (red line) from the period 15/11/1975 - 18/4/2014 and X-class flares (green dashed line) from the period 1/1/1976 – 5/5/2015 as a function of **Venus** heliocentric longitude. BIN=12°.

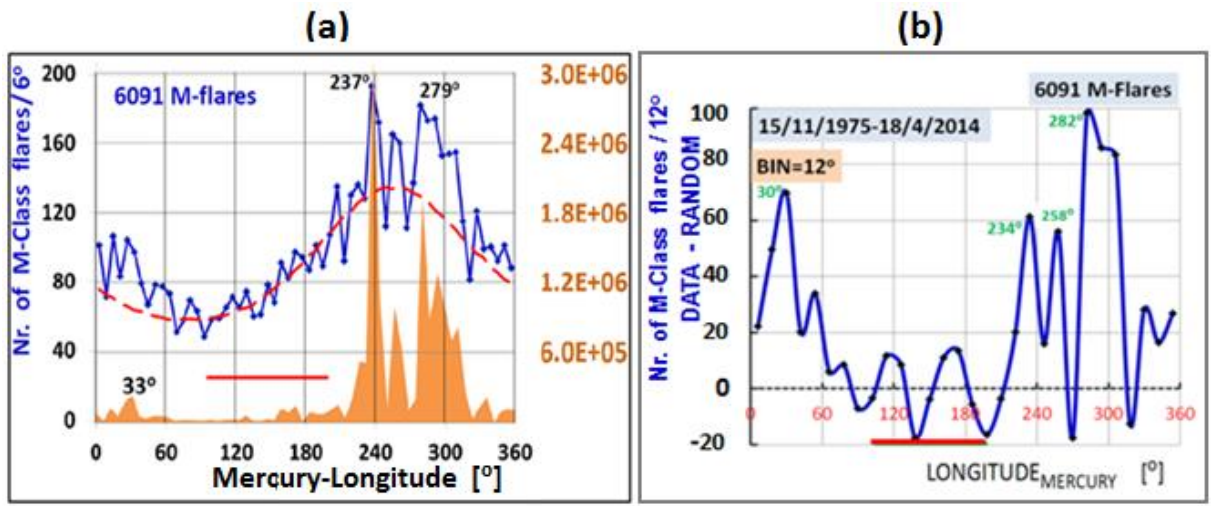

**Figure 4.** **(a)** Number of M-class flares as a function of **Mercury** heliocentric longitude (blue line) of the 6091 M-class Flares. The dashed red line is the normalised stay time simulation due to Mercury orbital eccentricity, which a random flare occurence would follow. The "multiplication spectrum" of the partial spectra of the 4 solar cycles is shown in orange. BIN=6°. **(b)** The residual spectrum, data (blue line) minus the isotropic M-flare occurrence (red dashed line) on the *left*, cancels the eccentricity related modulation. BIN=12°.

Excess peak candidates in the raw data (*left*), become clearly visible in the "multiplication spectrum" (orange, on the *left*) and the residual spectrum (*right*). For comparison, the null hypothesis, i.e., no planetary correlation, should give statistical fluctuations around the zero line (*right*). The horizontal red bars show the region used for normalization.

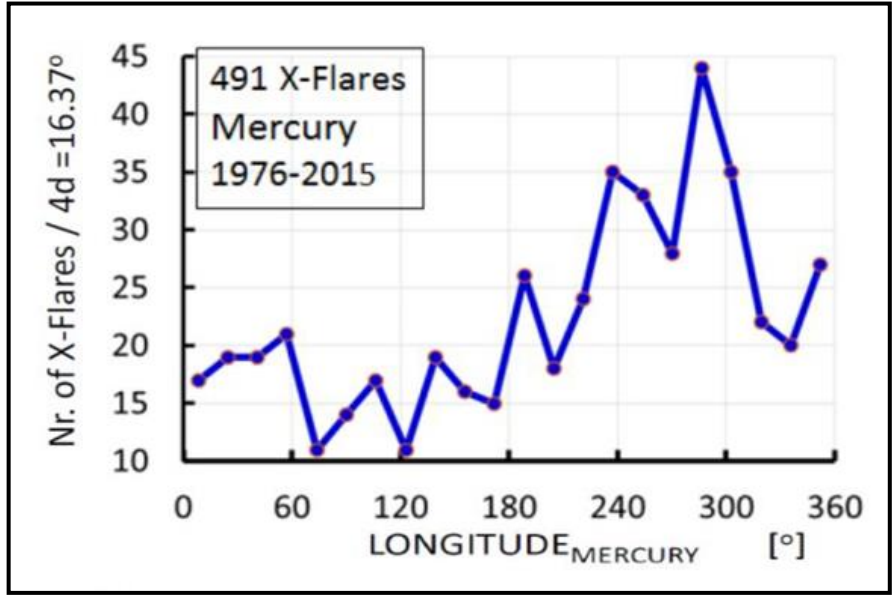

**Figure 5.**  Number of X-class Flares  from the period 1/1/1976 - 5/5/2015 as a function of the **Mercury** heliocentric longitude. The spectral shape resembles that of M-class Flares (see Figure 4a).

involvement in the workings of the Sun. Note that the excess above isotropic emission at ~292° is ~30%.

Throughout this work, for the flaring Sun, the statistical significance is based on the M-Flares sample using the raw data exclusively. Section 4.4 also gives the estimated statistical significance (assuming Poisson) of excesses with respect to a randomly active Sun. In fact, in these one-planet longitudinal distributions, six significant peaks are observed: Earth (2), Mercury (3) and Venus (1). Note that these estimates are conservative, since the appearance of one or more large peaks in one spectrum increases the mean value, resulting in a decrease of the calculated significance. An excess above randomly occurring event rate (red dashed line in Figure 4) with significance above 5σ is visible around 237° and 279°. Interestingly, taking into account the required slow speeds for gravitational lensing to occur, peaks are observed near the same heliocentric longitudes also in the spectra of Earth and Venus (see Figures 2 and 3).

In addition, in order to follow-up the behaviour in time of individual clustering candidates, we have divided the time series of X- and M-class flares in 4 sub-periods defined by the 4 solar cycles. Then we have calculated the product between each bin of the 4 partial spectra, which we call "multiplication spectrum". Analytically, the "multiplication spectrum" is given by

$$Y_{TOT}(J) = \Theta_1(J) \cdot \Theta_2(J) \cdot \Theta_3(J) \cdot \Theta_4(J),$$

where the subscript denotes the solar cycle: 1 (1975-1986), 2 (1986-1997), 3 (1997-2009) and 4 (2009-2014) and J denotes the bin number. We used three different bin widths (6°, 12°, 16°) to describe a 360° circle. $Y_{TOT}(J)$ is equal to the multiplied total number of Flares per bin. TOT is used to designate whether the first 3 or all 4 solar cycles have been used to derive the "multiplication spectrum".

Figure 6 shows the "multiplication spectra" for the Earth. The peaks in this Figure are occurring around the same longitudes as in Figure 2, but they exhibit a much better signal-to-noise ratio. The same is true also for the other "multiplication spectra" (see Figures 4,7,8). For Mercury, the "multiplication spectrum" for M-Flares is overlaid in orange in Figure 4a.

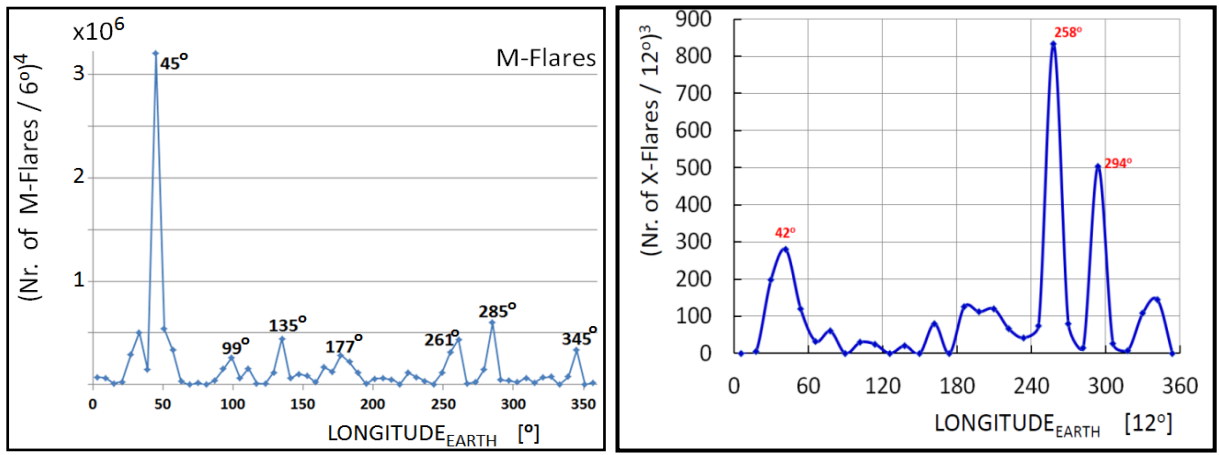

**Figure 6.** The multiplication of the number of M-flares (left) and X-flares (right) occurring in different solar cycles is shown as a function of the **Earth** heliocentric longitude in bins of 6° for M-flares and 12° for X-flares. Only those X-flares from the first 3 solar cycles are used due to weaker statistics. Note, the "exponent" on the Y-coordinate designates whether 4 or 3 solar cycles have been used in deriving the "multiplication spectrum".

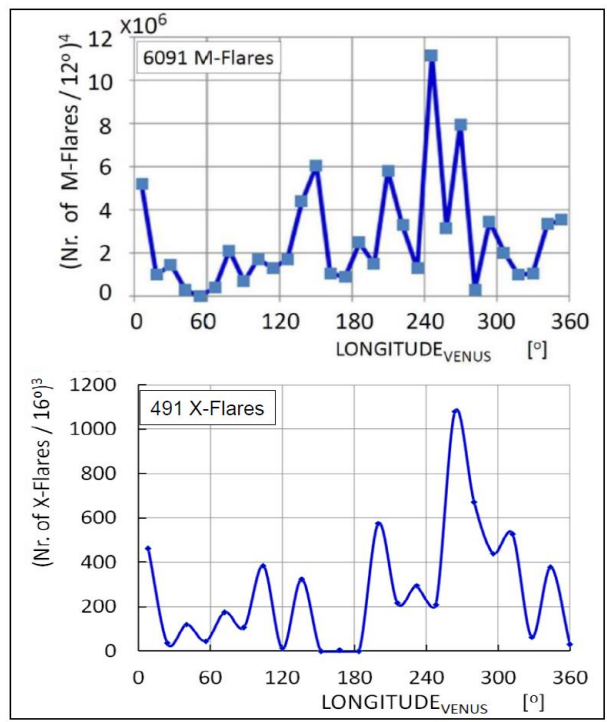

**Figure 7.** The multiplication of the number of M- and X-flares occurring in different solar cycles is shown as a function of the **Venus** heliocentric longitude in bins of 12° for M-flares and 16° for X-flares. X-flares are used only from the first 3 solar cycles due to weaker statistics. Note, the "exponent" on the Y-coordinate designates whether 4 or 3 solar cycles have been used in deriving the "multiplication spectrum".

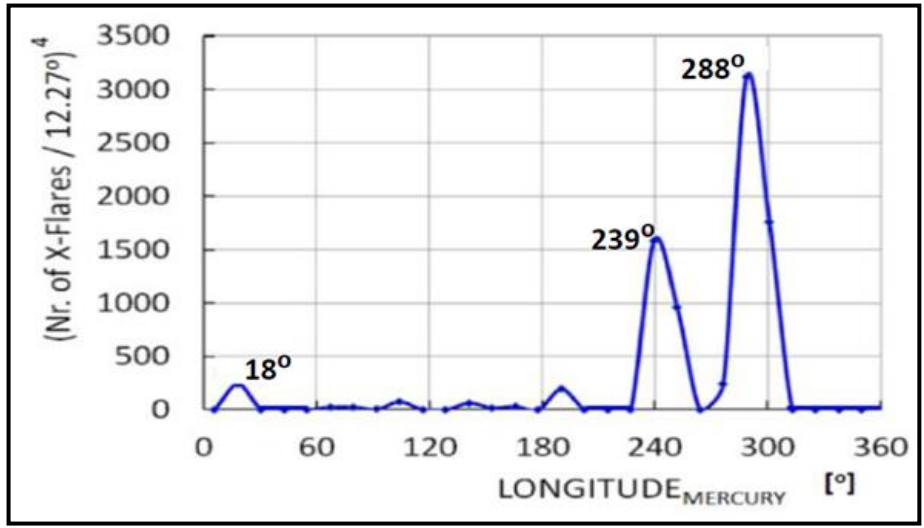

**Figure 8.** The "multiplication spectrum" for **Mercury** of X-class Flares for the 4 solar cycles. The excess at the double peaked region around 230° to 280° is apparent (for comparison, see Figure 5). Note, the "exponent" on the Y-coordinate designates that all 4 solar cycles have been used in deriving the "multiplication spectrum".

So far, our analysis has considered the effect of each planet regardless of the position of the other planets. The next logical step, discussed here in the case of Mercury and Venus, is to look whether combining the effect of the two planets has an influence on the observed distributions. In order to prove this hypothesis, we have plotted again the number of M-flares as a function of Mercury longitude adding the constraints of Venus being within ±60° around 260°, which defines two opposite sectors around the angular region where we have observed a large excess of flares. Figure 9 shows the spectra derived with these constraints. The difference among the number of M-class flares in the upper and lower distributions in Figure 9 is ~36%, while the expected isotropic distributions (blue lines in Figure 9), averaged over almost 4 solar cycles are within 1.5% equal. By comparing the upper spectrum with the lower one in Figure 9 (middle column), we note:
1) The appearance of 3 narrow peaks (duration <2 weeks!) only in the Mercury-Venus configuration of the upper spectrum, occurring near the same longitudes of the unconstrained spectrum (Figure 4) with much higher significance.
2) The statistical significance of the integrated excess between both Venus positions, i.e., the difference of the number of M-flares between the two Venus sectors (2312 − 1701 = 611) has a 9σ statistical significance. Along with 1), this supports the driving idea of slow speed invisible stream(s) towards the Sun being at work.

### 4.2  Solar EUV-irradiance and combined planet trigger

In order to check our hypothesis with an independent observation, we have analyzed the solar emission in the EUV (Figure 1). The solar EUV emission is related to the solar corona mystery, where large discrepancies are observed at high energies ($E_\gamma$ >20 eV) of the solar spectrum compared to the expected behaviour of a blackbody at 5800K. Daily observations of solar EUV started on 1/1/1996, but almost uninterrupted high precision measurements, which are used in this work, cover the period from 3/2/1999 to 25/10/2015 [12]. It is worth noting that this period includes the deepest solar minimum (2008/2009) of the last two centuries [15].

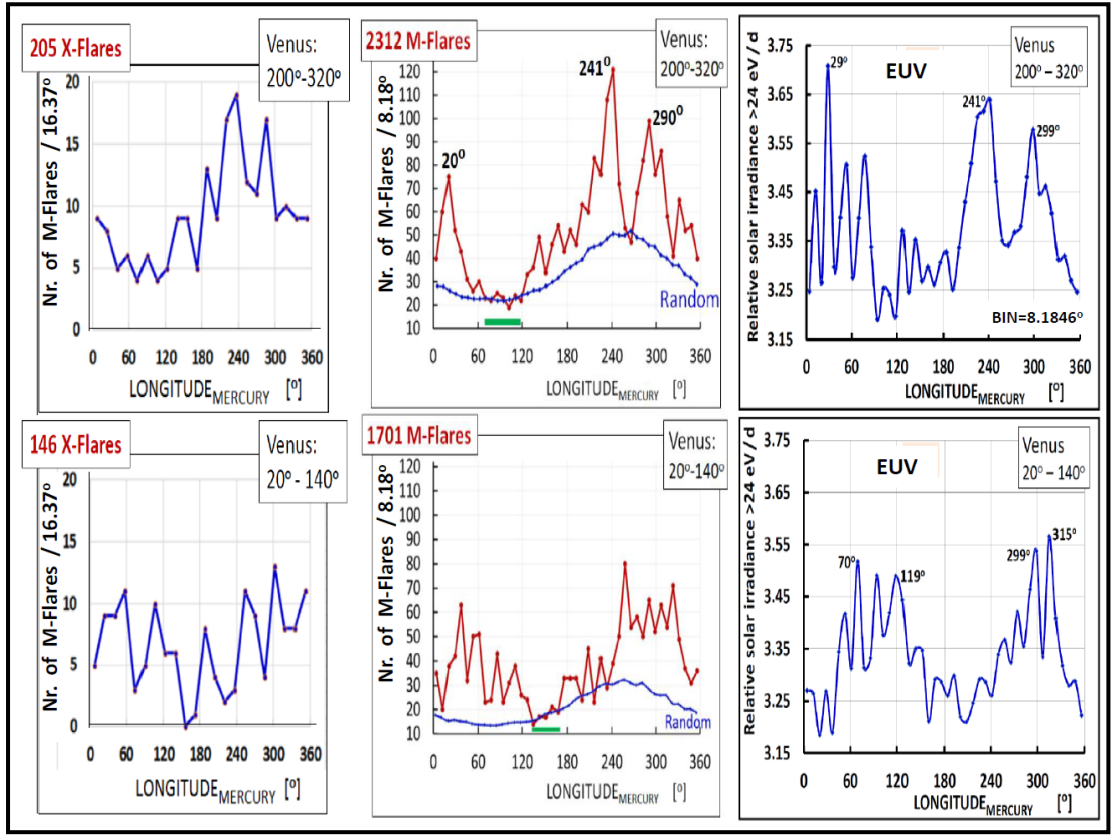

**Figure 9.** (UP: from left to right). Distributions of X-class Flares, M-class Flares and full disk EUV irradiance as a function of **Mercury** heliocentric longitude with the constraint of Venus being at longitude between 200°-320°. (DOWN) The same plots for Venus being between 20° and 140°. The smooth blue lines represent the expected normalized number of flares if equally distributed in time due to the large eccentricity of Mercury. The green bar shows the region used for normalisation. Note that the scale of each pair of plots is the same. For Poisson statistics, the difference in the number of M-Flares associated with the two different longitudinal positions of Venus (2312-1701=611) has a >9σ statistical significance. For X-flares the observed difference between the two Venus positions is >30%, although at the ~3σ level, due to 12x fewer X-flares.

Figure 9 shows the solar irradiance in the EUV (>24 eV) as a function of Mercury longitude, with the additional constraints that Venus should be within a heliocentric arc of 200°-320° or 20°-140°. This corresponds in the first case to Venus staying in a wide region of 120° around the preferred directions seen in the analyses of single planets (~240° to 300°), while the second region is 180° apart.

Also in these cases, the two Venus constraints result in largely different spectra, emphasizing the strong influence of the relative position of Venus and Mercury. Note that the same vertical scale is used for the corresponding UP and DOWN plots in Figure 9. Even more remarkably, one observes that the M-Flares and the EUV spectra (Figure 9 (UP)), peak at the same longitudes of ~241° and ~290°. We note here that the longitudinal position of Venus in the upper spectra of Figure 9 is centered around (200°-320°), which is the direction of the Galactic Center (266°).

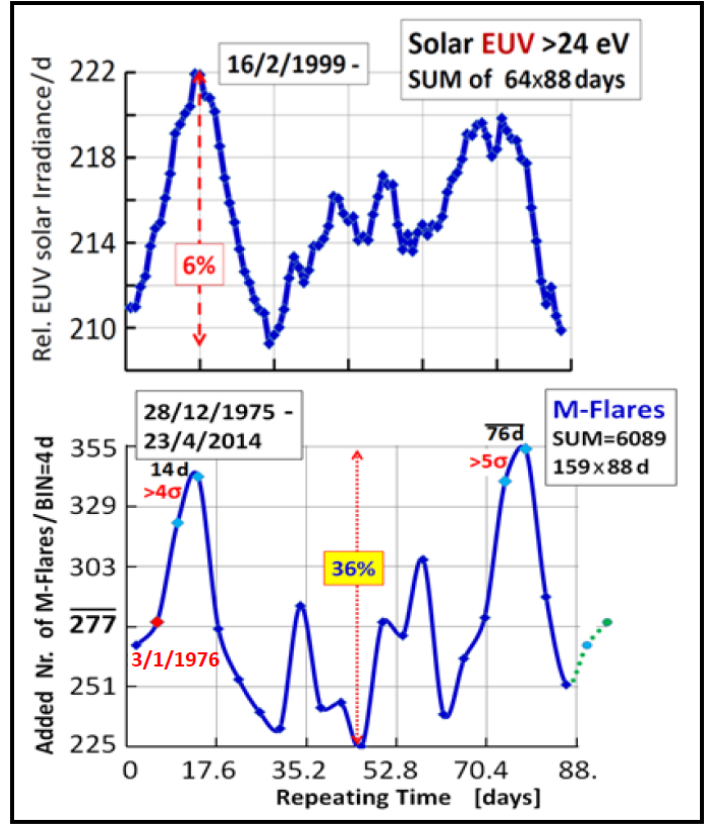

**Figure 10.** The sum of 88 consecutive days for solar EUV (UP) and M-Flares (BOTTOM). The fluctuations of the upper plot show that the measured EUV values are accurate below the ~1% level, which is sufficient for the purpose of this work. The zero point in the X-coordinate coincides (within 1 bin) with Mercury's heliocentric longitude Θ=0°. The BIN size is equal to 1 day and 4 days, for EUV and M-flares, respectively.

### 4.3 The non-random active Sun

From the Fourier spectra of the active Sun, like the ~154 days gamma rays Rieger flare periodicity discovered in 1984 [16], it is accepted that the solar activity is not random. Using the same data of M-flares and EUV irradiance of this work, we show below an alternative method to exclude Poisson statistics for these phenomena. In order to do that, we have divided the data in consecutive periods of 88 days (which corresponds to the Mercury orbital period of 87.969 days). We have then added all periods together day per day (Figure 10). If the data would be randomly distributed, one should expect a flat distribution, since our procedure corresponds to averaging 159x and 64x the daily number of flares or the EUV emission strength, respectively. Instead, both distributions show at least two striking peaks within the 88 days, confirming the non-Poissonian character of the two solar activities (see section 4.4). One should also note that by plotting quantities as a function of the day, we completely remove any eccentricity related effect.

## 4.4 Signal significance

The single planetary heliocentric longitudinal distributions of the flaring Sun (Figure 2-4) show various peaks which are beyond that expected from an isotropic active Sun. Assuming Poisson statistics, the observed excess is significant (>5σ) compared to a randomly occurring rate. The Earth's longitudinal distribution (Figure 2) gives peaks above the mean value (101.6±1.3), at: Θ≈45° (>6σ), Θ≈255° (5.2σ), Θ≈177° (4.8σ), Θ≈291° (4.7σ), Θ≈135° (3.2σ). Similarly, the longitudinal distribution for Venus shows two peaks above the mean value (203±2.6) at Θ≈258° (5.3σ) and Θ≈145° (3.2σ). The corresponding spectrum for Mercury shows 3 peaks above the isotropic distribution (dashed line in Figure 4a); the peak positions are at Θ≈292° (6σ), Θ≈240° (5.1σ), Θ≈33° (~5σ). This excess becomes more visible in Figure 4b, which removes the large eccentricity effect by subtracting those simulated from the raw data (assuming a randomly flaring Sun).

The significance of planetary correlations becomes even stronger, when we consider the combined effect of two planets, i.e., Mercury and Venus (Figure 9). Along with the three narrow peaks, the total number of M-flares for the one position of Venus (200°-320°) is by ~36% above that of the opposite Venus position (20°-140°). In addition, if the Venus position interval is narrowed from 120° to 90°, then the excess increases from 36% to 47% (not shown here). In either case, the statistical significance following Poisson is at the 9σ level!

The two peaks in the EUV (Figure 10) are of highest statistical significance. Assuming Poisson statistics, the >4σ and >5σ significance (Figure 10 BOTTOM) of the peaks of the flaring Sun at day 14 and 76, respectively, are underestimated, since they refer to the mean value (=277 cts) derived from the whole spectrum.

## 5. THE EARTH ATMOSPHERE: measurements of the dynamic ionosphere

### 5.1 Planetary correlations

We applied the same concept to the analysis of the total electron content (TEC) of the Earths atmosphere. While the global electron content of the ionosphere depends primarily on the variable solar EUV irradiance, its variations show anomalies, which have not been understood so far [17]. For example, the measured TEC during the December solstice exceeds that around the June solstice by about 20%, which cannot be explained by the annual solar irradiance modulation due to the varying Sun-Earth distance. We investigated the possibility that these anomalies could be connected to planetary lensing, using the same hypothesis, but this time using the Earth as the target.

We present here the results for the time dependent global electron content of the ionosphere (Figure 11), derived from an uninterrupted sequence of 6573 Global Positioning System (GPS) daily averaged measurements of TEC, from 1/1/1995 to 31/12/2012 [12,18]. This period includes the extremely deep solar minimum between 2008 and 2009 [15], which induced a quieter behaviour of the electron content of the ionosphere.

Figure 12 shows the daily total electron content as a function of the heliocentric longitude of the Earth, with no constraint applied. The ~20% electron content variation at the winter and summer solstices is clearly visible, and so far this variation has no accepted explanation. We then repeat the same procedure as used for the M-flares and EUV, by looking for possible correlations with the position of Mercury and Venus. The main results of this analysis are shown in Figure 13. The two columns show the heliocentric longitudinal distribution of Mercury and Venus, requiring a second planet (among the 3 inner planets) to be in a wide heliocentric longitudinal range 20°-140° *vs*. 200°-320°, which are symmetrically 180° apart. Three plots refer to the solar maximum period (1997-2006), while a similar comparison is shown for the extreme solar minimum period (column A, DOWN). Since the derived rates are normalized to 1 day, longitudinal modulations due to planetary eccentricity are factored out. Considering long observation time periods of 10 or even 18 years, the derived distributions should be rather isotropic. This is, however, in contrast with observed amplitude differences at the 20-40% level, showing also narrow peaks, which appear more pronounced in Figure 13 (B).

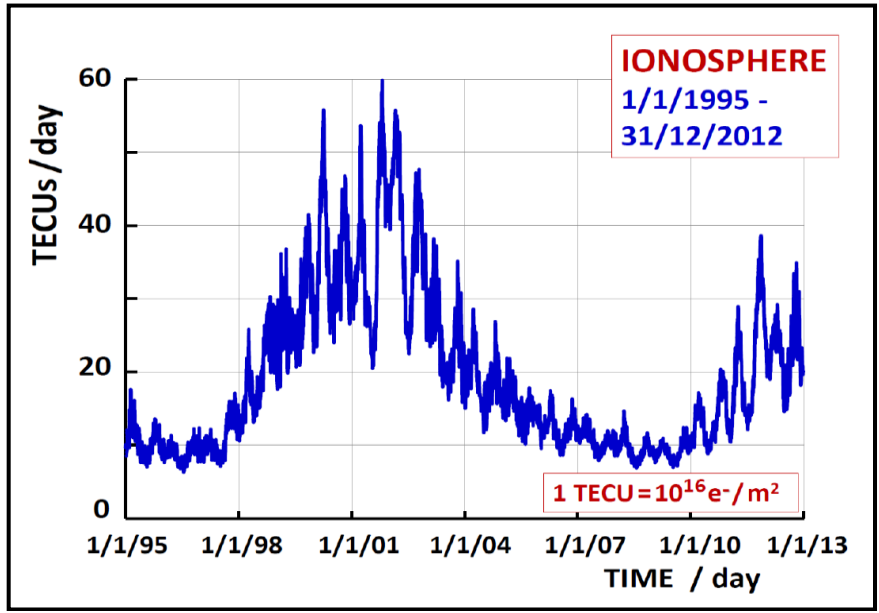

**Figure 11.** The time series of the measured total electron content (TEC) of the Earth's ionosphere averaged over one day during the 18 year period 1995 – 2012 [12]. During this period the Moon performed 223 geocentric orbits.

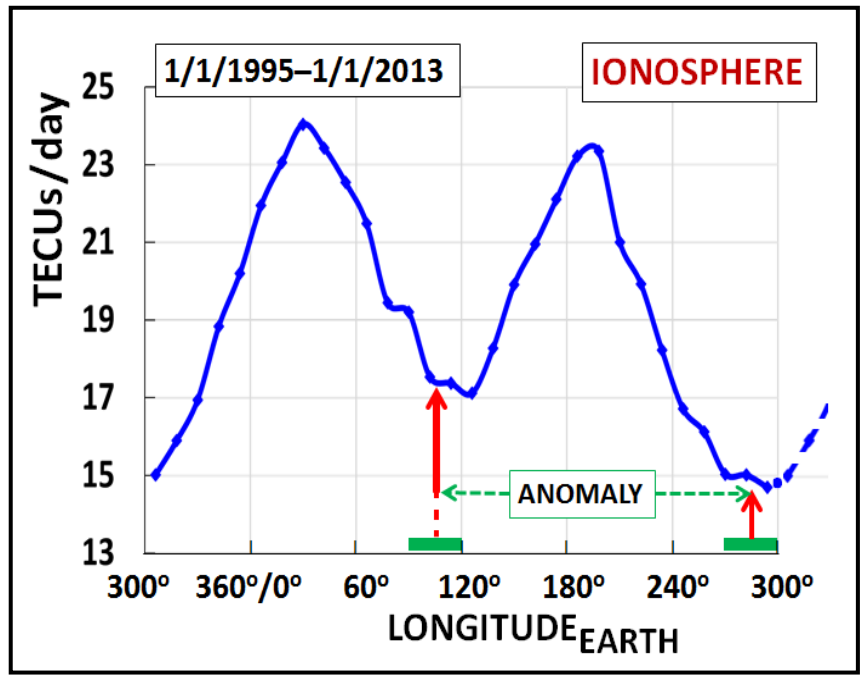

**Figure 12.** The measured atmospheric total electron content [in TECUs, 1 TECU= $10^{16}$ e/m$^2$] as a function of the **Earths** heliocentric longitude averaged over one day (1995-2012) [12]. The zero point of the X-coordinate is shifted to the right, in order to show both broad peaks uninterrupted. The green bars cover the longitude segments 90°-120° and 270°-300°, which are used in the analysis (see Figures 14, 15). The short dashed line on the right repeats the beginning of this curve, in order to make the minimum around 300° more visible. The difference in rate between the two minima is about 20% and reflects the ionosphere's "annual anomaly" [17] between the two solstices in December and June.

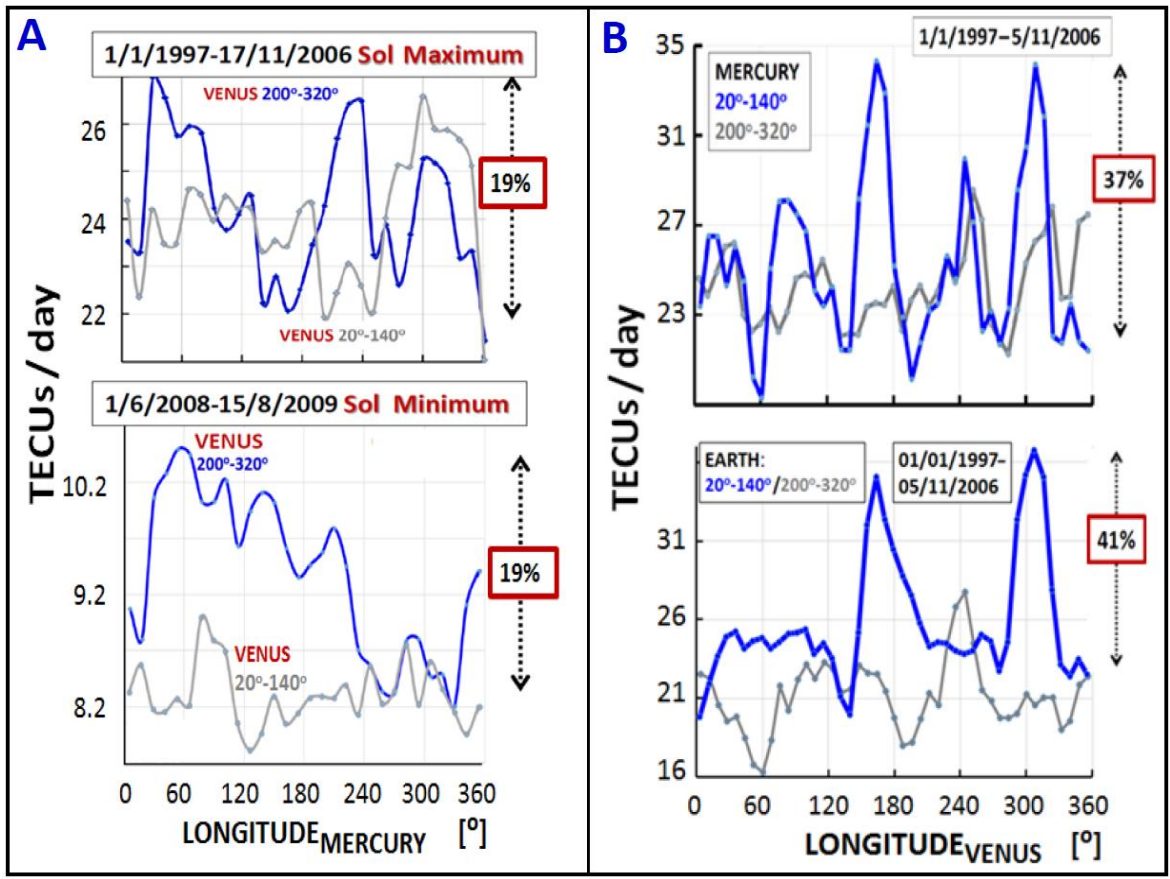

**Figure 13.** The daily measured longitudinal distributions of Earth's atmospheric total electron content [12] for different planetary configurations and time periods of the solar cycle: (A) The total electron content [TECUs] as a function of **Mercury** heliocentric longitude during the solar maximum period 1997-2006 (UP) and during theextremely deep solar minimum 2008-2009 (DOWN). The thick blue and the thin grey lines are associated with Venus being in one of the two opposite 120° wide orbital arcs. (B) TECUs as a function of the **Venus** longitude during the same solar maximum requiring Mercury (UP) and Earth (DOWN) to be in one of the 120° segments. Note that the two strongest peaks (in blue) appear in both.

## 5.2 Luna correlations

We have also studied the TEC as a function of the orbital position of the Moon around the Earth, using only the periods around the solstices, in a sector of 30° around the minima as is indicated in Figure 12. Figure 14 (*left*) shows the TEC distributions of these data as a function of the Moon phase. Note that during the Earth's propagation by 30° in longitude, the Moon completes one geocentric orbit. Taking into account the 223 orbits the Moon performed around the Earth during the 18 years described by the dataset, randomly occurring TEC excursions should average out, or at least both distributions in Figure 14 (*left*) should have a similar shape. Figure 14 (*right*) shows the difference between the two distributions in Figure 14 (*left*), exhibiting a variation of a factor 6 between maximum and minimum: remarkably, the position of the maximum coincides with New Moon.

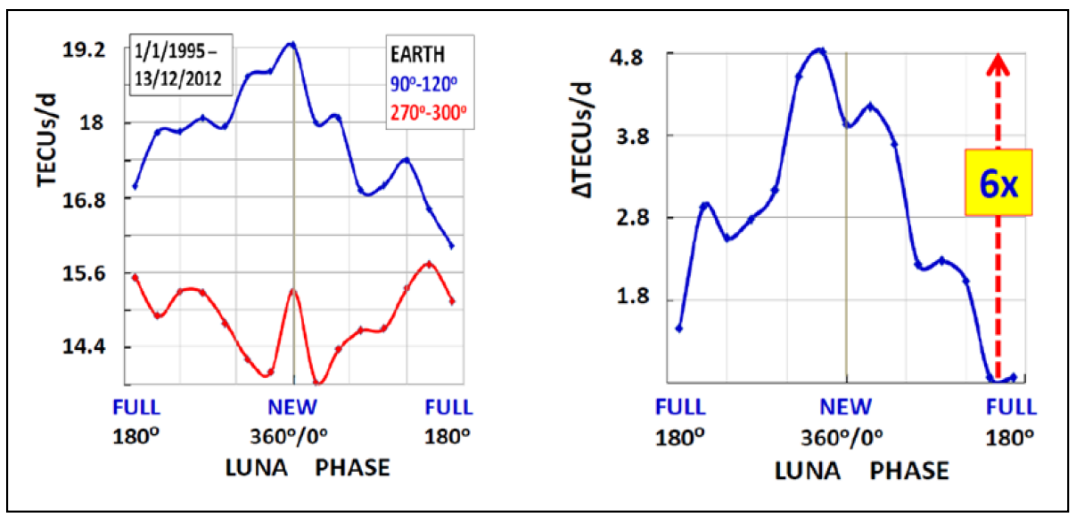

**Figure 14.** TECUs as a function of the **Moon Phase**, while the Earth is in one of the two 30° orbital segments around the solstices (*left*) and the difference between the winter-summer solstices (*right*) (see Figure 12).

This correlation (Figure 14) fits the assumption of this work, which assumes a massive stream of invisible matter coming from the direction of the Sun around December, which is gravitationally focused at the Earth by the combined effect of the Sun and the interposed Moon. We note that for an Earth observer, only during December the galactic center is aligned (within ~5.5°) with New Moon – Sun.

Finally, Figure 15 shows the distributions of the TEC as a function of the Earth heliocentric longitude, with the constraint of the Moon being (within 30°) in one of the four different lunar phases. Differences in the distributions are clearly seen around Full Moon and New Moon. The relative height of the two maxima (Figure 15 UP)*,* depends on the position of the Moon being consistent with the results of Figure 14, where the role of the Moon-Earth are interchanged.

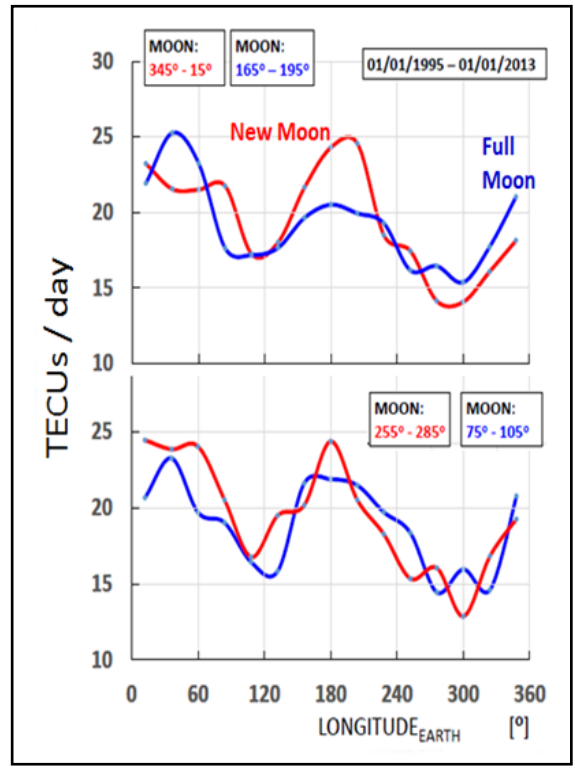

**Figure 15.** The daily measured atmospheric total electron content [12] as a function of Earth's heliocentric longitude during the period 1995-2012 for different 90° orbital segments of the Moon: (UP) during New Moon (red) and Full Moon (blue). (DOWN) when the Moon is sideways to the Sun-Earth direction, 75°-105° and 255°-285°, with New Moon being equal to 0° in an anticlockwise geocentric reference frame.

## 6. DISCUSSION

By analyzing the occurrence of X- and M-flares during the last 4 solar cycles and the full disk EUV irradiance of the Sun, we find strong evidence that the occurrence of these phenomena is strongly modulated by the position of Earth, Venus and Mercury around their heliocentric longitude. A preferred direction around 270° is common to all three planets, when their lensing effect is studied independently. The effect is further enhanced, as we show for the case of Mercury and Venus, when one takes into account the relative position of the two planets. This observation supports our working hypothesis that the activity of the Sun is triggered by the influx of invisible massive matter and that this matter has some preferred direction or stream, which gets gravitationally lensed by the planets.

In addition, averaged daily GPS TEC measurements of the Earth ionosphere also show a marked planetary correlation, supporting our lensing scenario, which is further reinforced by the observation of the effect of the position of the Moon in modulating the TEC of the ionosphere.

We note that the anomalously high electron content of the ionosphere in December coincides with the alignment Galactic Center - Sun - Earth, i.e. the same direction around 270° previously described for the active Sun and that this effect becomes even more enhanced when the Moon is aligned around the same direction and interposed between the Sun and the Earth.

In this work, the observations of
a) multiple orbital cycles (~40 for the Earth, ~65 for the Venus and ~160 for the Mercury),
b) single and double planetary dependence of the flaring Sun,
c) the strikingly similar EUV planetary dependence, based on data taken from the satellite SOHO on a totally different orbit, and
d) the planetary and lunar dependence of the Earth's degree of ionization,

make the existence of an *unidentified* event selection bias highly improbable, since the three datasets are independent and refer to different phenomena.

The results of our analysis support our working hypothesis of streams of unidentified massive particles gravitationally focused by the planets of the solar system. The identification of the nature of the predicted slow moving stream(s) is left for future work: one has to suppose that this invisible massive matter has some form of interaction with normal matter, but at this point we cannot infer any useful indication of its nature.

In this work we have demonstrated the non-Poissonian behaviour of planetary distributions associated with solar and ionospheric activity. The statistical significant excess in Figure 9 is associated with a short solar activity excursion (FWHM ≈ 10 days). The gravitational lensing scenario allows for the observed short time correlations. This result excludes planetary tidal-force inspired models, which by their nature extend smoothly over an orbital period.

In future we will elaborate more refined analyses introducing more planets and more constraints as well as new datasets, in the attempt to determine the direction of the stream(s) and the velocity spectrum of their constituents. Also, the (re)-analysis of the experiments searching for dark matter following our approach might give unexpected results.

Interestingly, very recently, an independent analysis [19] has confirmed our results with M-Flares.

**Acknowledgments**

In the course of this work we have profitted a lot from the various kinds of support / feedback we had from Vassilis Anastassopoulos, Tullio Basaglia, Giovanni Cantatore, Patla Biju, Yiding Chen, Partha Chowdhury, Leonid Didkovsky, Sue Foffano, Anne Gentil-Beccot, Nectaria Gizani, Bettina Hamoudi, Klemens Hocke, Rajmal Jain, Eleni Petrakou, Thomas Papaevangelou, Salomé Rohr, Yannis Semertzidis, Stan Solomon, Igor Tkachev, Ioannis Tsagris, Astrid Veronig. In particular we are thankful to Astrid Veronig for providing us with the list of M- and X-flares, and to Leonid Didkovsky for the list of the atmospheric electron data (1995-2012) as well as the EUV data (since 1996) from the CELIAS/SEM experiment on the Solar Heliospheric Observatory spacecraft (SOHO is a joint ESA and NASA mission). We also thank Yannis Semertzidis for reading the first draft and providing very useful suggestions, and Wolfgang Funk for his encouraging interest in this work. We also thank the anonymous referees for their careful reading and many suggestions, which helped improve the original manuscript.

## REFERENCES

[1] C.A.J. O'Hare, A.M. Green, Phys. Rev. D90 (**2014**) 123511, and references therein; arXiv:1410.2749v2 .

[2] J.J. Fan, A. Katz, L. Randall, M. Reece, A Dark-Disk Universe, Phys. Rev. Lett. 110 (**2013**) 211302 arXiv:1303.3271v2, and Double-Disk Dark Matter Phys. Dark Univ. 2 (**2013**) 139 arXiv:1303.1521v2

[2] S. Wedemeyer *et al.*, Solar Science with ALMA, arXiv:1504.06887v2 (**2015**).

[3] V. Polito, J.W. Reep, K.K. Reeves, P.J.A. Simões, J. Dudík, G. Del Zanna, H.E. Mason, L. Golub, arXiv:1512.06378 [astro-ph.SR] (**2015**).

[4] H.P. Warren, D.H. Brooks, G.A. Doschek, U. Feldman, arXiv:1512.04447 [astro-ph.SR] **(2015).**

[5] V. Hansteen, B. De Pontieu, M. Carlsson, J. Lemen, A. Title, P. Boerner, N. Hurlburt, T.D. Tarbell, J.P. Wuelser, T.M.D. Pereira, E.E. De Luca, L. Golub, S. McKillop, K. Reeves, S. Saar, P. Testa, H. Tian, C. Kankelborg, S. Jaeggli, L. Kleint, J. Martínez-Sykora, Science, 346 (**2014**) 1255757. DOI: 10.1126/science.1255757 / arXiv:1412.3611v1 [astro-ph.SR].

[6] R. Lionello, C.E. Alexander, A.R. Winebarger, J.A. Linker, Z. Mikić, arXiv:1512.06146 [astro-ph.SR] (**2015**). http://xxx.lanl.gov/abs/1512.06146

[7] B.R. Patla, R.J. Nemiroff, D.H.H. Hoffmann, K. Zioutas, ApJ. 780 (**2014**) 158, and references therein; doi:10.1088/0004-637X/780/2/158.

[8] B.R. Safdi, M. Lisanti, J. Spitz, J.A. Formaggio, Phys. Rev. D90 (**2014**) 043001; DOI: 10.1103/PhysRevD.90.043001 / arXiv:1404.0680v1.

[9] E. Möbius P. Bochsler, M. Bzowski, D. Heirtzler, M.A. Kubiak, H. Kucharek, M.A. Lee, T. Leonard, N.A. Schwadron, X. Wu, S.A. Fuselier, G. Crew, D.J. McComas, L. Petersen, L. Saul, D. Valovcin, R. Vanderspek, P. Wurz, ApJS. 198 (**2012**) 11; http://iopscience.iop.org/0067-0049/198/2/11/.

[10] NASA:http://www.jpl.nasa.gov/copyrights.php; http://www.caltech.edu/copyright/

[11] Astrid Veronig, private communication (**2014**).

[12] Leonid Didkovsky, private communication (**2014**).

[13] See, e.g., M.J. Aschwanden, S.L. Freeland, ApJ. 754 (**2012**) 112, https://arxiv.org/abs/1205.6712 .

[14] D.F. Ryan, R.O. Milligan, P.T. Gallagher, B.R. Dennis, A.K. Tolbert, R.A. Schwartz, C.A. Young, ApJSS. 202 (**2012**) 11, arXiv:1206.1005v2.

[15] S.C. Boscardin, C. Sigismondi, J.L. Penna, V. D'Avila, E. Reis-Neto, A.H. Andrei, arXiv:1512.03358 [astro-ph.IM] (**2015**).

[16] E. Rieger, G.H. Share, O.J. Forrest, G. Kanabach, C. Reppin, E.I. Chupp, A 154 day periodicity in the occurrence of hard solar flares? Nature 112 (**1984**) 623.

[17] Y. Chen, L. Liu, H. Le, W. Wan, Earth, Planets and Space 66 (**2014**) 52, http://www.earth-planets-space.com/content/66/1/52; see also M. Mendilloa, C.-L. Huanga , X. Pib , H. Rishbetha, R. Meier, J. Atm. Sol.-Terr. Phys. 67 (**2005**) 1377 (doi:10.1016/j.jastp.2005.06.021).

[18] L. Didkovsky, S. Wieman, J. Geophys. Res. Space Physics, 119 (**2014**) 4175 (doi:10.1002/2014JA019977 ).

[19] Horst Fischer / University of Freiburg-Germany, private communication [**2017**].